\documentclass[12pt,preprint]{aastex}

\received{}
\revised{}
\accepted{}
\ccc{}
\cpright{}{}

\newcommand{\dd}{\delta}

\newcommand{\Th}{\Theta}
\newcommand{\lra}{\longrightarrow}

\newcommand{\unitt}{{\bf \hat e_\Th}}

\newcommand{\tensor}{{\baselineskip=2pt
\vbox{\hbox{$\leftrightarrow$}\vglue .02mm
\hbox{$\bf H$}}}}

\newcommand{\gggg}{$\gamma \gamma \lra e^- e^+$\/}
\begin{document}

\title{COLLIMATED ESCAPING VORTICAL POLAR $\bf e^-e^+$ JETS 
INTRINSICALLY PRODUCED BY ROTATING BLACK HOLES AND PENROSE
PROCESSES}

\author{Reva Kay Williams}

\affil{Department of Astronomy, University of Florida, 
Gainesville, FL 32611}

\email{revak@astro.ufl.edu}

\begin{abstract}
In this paper, I present results from theoretical and numerical 
(Monte Carlo) {\it N-particle\/} fully relativistic 4-D analysis of
Penrose scattering processes (Compton and \gggg) in the 
ergosphere of a supermassive or stellar mass Kerr (rotating)
black hole.  Specifically, the escape conditions and the escaping
orbits of the 
Penrose pair production (\gggg)  electrons are analyzed, 
revealing 
that these particles escape along collimated, jet geodesic trajectories 
encircling the polar
axis.  Such collimated vortical tightly wound coil-like  trajectories
 of relativistic particles 
are inherent properties
of rotating black holes.
The helical polar angles of escape for these $e^-e^+$ pairs 
range from $\sim 40^o$ to $\sim 0^o.5$ (for the highest energy
particles).   These jet
distributions appear to be  consistent with the astrophysical jets
of active galactic nuclei (AGNs) and galactic black holes,
and suggest a mechanism for precollimation within the
inner radius of the dynamically stable accretion disk. 
\end{abstract}

\section{Introduction}
\label{sec:intro}

We now have observational evidence that black holes indeed exist in
nature.  They are at the cores of quasars and other active galactic
nuclei (AGNs) as well as sources in our Galaxy, commonly referred
to as microquasars or galactic black holes.
Many of these sources are associated with polar jets emanating
from their cores. 
Black holes were theoretically predicted from Einstein's theory 
of general relativity.  Theoretical and numerical calculations 
(Williams 1991, 1995, 1999,  2001,  2002a, 2002b, 2003),
described briefly below,
show that Penrose (1969) gravitational-particle scattering 
processes are
sufficient to describe energy-momentum extraction from a rotating
Kerr (1963) black hole, from radii within the marginal stable orbit
($r\la r_{\rm ms}\simeq 1.2M$, in gravitational units: $c=G=1$;
${a/M}=0.998$, where $a$ is the angular momentum per unit 
mass parameter and $M$
is the mass of the black hole), 
while electromagnetic interactions
or magnetohydrodynamics (MHD) appear to govern the ``flow'' 
of polar jets of the extracted
particles, escaping away from the central source, out to the 
observed distances,
as suggested by observations (Junor, Biretta, \& Livio 1999).

In the primary paper cited above (Williams 1995),
theoretical model calculations involving
Monte Carlo computer simulations
of Compton scattering and electron-positron ($e^- e^+$)
pair production processes in
the ergosphere  of a supermassive ($\sim 10^8 M_\odot$)
rotating black hole are presented.
Particles from an accretion disk surrounding the rotating black hole
fall into
the ergosphere and scatter off particles that are in bound equatorially
and nonequatorially confined orbits. The accretion flow is assumed to
be of the form of the so-called
{\it two-temperature} bistable
disk model, in which the disk can in principle exist in two phases:
a thin disk (Novikov \& Thorne 1973) and/or ion torus (Lightman \& Eardley 
1974; Shapiro,
Lightman, \& Eardley 1976; Eilek 1980; Eilek \& Kafatos 1983), where
the electrons and protons (or ions) can have separate temperatures of 
up to $\sim 10^9$~K and $\sim 10^{12}$~K, respectively.
Note that disk models of this sort
are also
referred to as thin disk/ion corona models, and
more recently,
the ion corona has been called an advection dominated
accretion flow (ADAF; Mahadevan, Narayan,  \&  Krolik 1997).
The Penrose mechanism, in general,  allows
rotational energy
of a Kerr black hole (KBH) to be extracted by
scattered particles escaping from the
ergosphere to large distances from the rotating black hole.
The results of
these model calculations
 show that the Penrose mechanism is capable of
producing the astronomically observed high energy particles
($\sim $~GeV) emitted by quasars and other
AGNs.  This mechanism can extract hard X-ray to $\gamma$-ray
photons from Penrose Compton scatterings of initially
low energy soft X-ray photons by target orbiting electrons in the
ergosphere.
The Penrose pair production (\gggg) processes         
allow  relativistic $e^- e^+$ pairs to escape with energies up to
$\sim 4$~GeV, or greater depending on the form of the
accretion disk; these pairs are produced when infalling low energy
photons collide with bound, highly blueshifted photons
at the $\it photon~ orbit$.
This process may very well be the origin of the relativistic
 electrons inferred  from observations to emerge from the cores of
AGNs. 
Importantly, these model calculations show that
the Lense-Thirring effect (Thirring \& Lense 1918), 
i.e., the dragging of 
inertial frames into rotation, 
inside the ergosphere, caused by the
 angular momentum of the rotating black hole,
results in a gravitomagnetic
force being exerted on
the scattered escaping particles. This force (which is the gravitational 
analog or resemblance
of a magnetic  force) produces
asymmetrical particle emissions in the
polar direction, above and below the equatorial plane,
consistent with the asymmetrical or one-sided jets
observed in radio strong AGNs (Williams 2002a, 1999)\footnote{Figures~1(c) 
and~1(d) of Williams (1999)
are incorrect.  Figures~4(e) and~4(h), respectively, of Williams (2002a)
are the correct figures, where in  Figure~4(h) the target electrons have
both positive and negative equal absolute value polar coordinate
angular momenta.}.   The
dragging of inertial frames also causes the Penrose
escaping particles to escape along vortical trajectories
(as discussed in the following paragraphs).

These Penrose processes can apply to any size rotating black hole
and, in general, to any type of relativistic elementary particle
scattering energy-momentum exchange  process, inside the ergosphere,
allowing particles to escape with rotational energy-momentum from the
KBH.  Even in the context of MHD, according to the guiding center
(Landau \& Lifshitz 1975) approximation, the single-particle approach
is essential close to the black hole (de Felice \&  Carlotto 1997;
Karas \& Dov\v{c}iak 1997; de Felice \& Zonotti 2000), i.e., the behavior
of individual particles moving along geodesics in the strong central
gravitational force field is also that of the bulk of fluid elements.
This suggests that even though a MHD simulation is not performed in
this present paper, the results will be valid for the trajectories of
particle flows and $e^-e^+$ pairs produced near the black-hole event
horizon.  Nevertheless, the Penrose analysis presented here should
be considered in the context of a  full-scale relativistic MHD
simulation.  However, because of the proximity of the Penrose processes
to the horizon, and the dominating effect of gravity,  the resulting
trajectories are expected to be approximately the same in the
MHD regime.

In the model calculations summarized above, in which energy-momentum
is extracted from a rotating black hole,
it is found
that particles escape with relativistic velocities along
 vortical trajectories, above and below
the equatorial plane, with small helical angles of escape, 
implying strong coil-like vortex plasma collimation (Williams  2001, 
 2002b).
Thus, from these model
calculations, it appears that the rotating black hole naturally produces
particle trajectories collimated about the axis of symmetry, i.e., the
polar axis.
Such vortical orbits or trajectories have been
discussed by some other authors
 (de Felice \& Calvani 1972;
de Felice \& Curir 1992;
de Felice \& Carlotto 1997;  de Felice \& Zanotti (2000);
see also Bi\v{c}\'{a}k, Semer\'{a}k, \& Hadrava 1993; 
Karas \& Dov\v{c}iak 1997).  Their independent findings,
deduced from the geodesic properties of the Kerr metric, and
referred to as geometry induced collimation,
can serve as confirmation of  Williams' (1991, 1995)
theoretical and numerical calculated results, or vice versa.  

In this paper, I examine the escape conditions and the 
resulting four-momentum vectors of the  Penrose pair production 
(\gggg) processes, showing 
that the particles indeed escape to infinity in the form of
vortical jets intrinsically collimated about the polar axis,
 because of the frame dragging of spacetime 
inside the ergosphere of the KBH. 
Importantly, we shall
see that these
$e^-e^+$ pairs escape without any appreciable interaction
with the dynamically stable accretion disk particles.
Now, although most of the Penrose Compton scattered photons escape
along vortical trajectories as well, we are concerned,
in this paper, only with the Penrose pair production
(\gggg) electrons, because
these are the particles
that compile the main constituents of the observed jets, being
responsible for the synchrotron radiation and Doppler
boosting (giving rise to superluminal motion). 
I refer the reader to the above references,
particularly Williams (1995), for a thorough
description of the ``Penrose-Williams'' (Williams 2002b) 
processes discussed in 
this present paper.

\section{Escape Conditions and Vortical Orbits}
\label{sec:orbits}

After the scattering events,  not all of the particles escape to 
infinity.  A set of escape conditions must be applied to see if
a particular particle escapes from the gravitational potential 
well of the black hole [see Williams (1995) and references therein
for further descriptions of the escape conditions].  It is known that 
outside 
the event horizon, 
the orbit of a photon (or an
unbound material particle with $E/ \mu_o > 1$), may 
have one or no
radial turning points for which $P_r\rightarrow 0$ 
(Piran \& Shaham 1977;
Williams 1995), where $E$ is the energy of the particle as measured
by an observer at infinity; $\mu_o $ is 
the rest mass 
energy; and  $P_r$ is the radial component of the 
covariant four-momentum vector 
$\bigl[P_\mu=\bigl(P_r,P_\Th,P_\Phi,-E \bigr)\bigr]$.  
Let  $L$ define the conserved azimuthal coordinate angular 
momentum ($=P_\Phi$), then if $E/L$
for that particle lies in the range
\begin{equation}
{E \over L} \leq {E_{\rm orb} \over L_{\rm orb}},
\label{eq:escape1}
\end{equation}
for a direct orbit not confined to the equatorial
plane [$Q>0$; $Q$ is the so-called Carter constant of motion
(Carter 1968):
\begin{equation}
Q=P_\Theta^2 + \cos^2\Theta \biggl[a^2 \biggl(\mu_o^2-E^2 \biggr)
  +{L^2 \over \sin^2 \Theta} \biggr],
\label{eq:carter}
\end{equation}  
where the value of $Q$ is zero for particles
whose motions are confined to the equatorial plane], 
there will be one turning point;
otherwise, there will be none, where $E_{\rm orb}$ is
the iso-energy orbit: the circular orbit of equal energy
at constant radius $r=r_{\rm orb}$, a potential turning point,
and $L_{\rm orb}$ is the 
corresponding azimuthal angular momentum.  
Another independent condition 
for the existence of a turning point for which $P_r\rightarrow 0$ is 
\begin{equation}
{Q \over E^2} \geq{Q_{\rm orb}
\over E_{\rm orb}^2},
\label{eq:escape2}
\end{equation}
where $Q_{\rm orb}$ is the corresponding $Q$ value of the iso-energy
orbit.
[Note, a bound circular orbit is considered as an orbit with a ``perpetual''
turning point (see Williams 1995).]  In equations~(\ref{eq:escape1}) 
and~(\ref{eq:escape2}), 
$E_{\rm orb}$ and $L_{\rm orb}$ are the conserved orbital energy
and azimuthal angular momentum as measured by an observer at infinity,
given by (Williams 1995)\footnote{Note the
typographic error in equation~(A17) of Williams (1995): $\mu_0$
is supposed to be
$\mu_0^2$.}
\begin{equation}
E=\left({r^2L^2+D+F\over G}\right)^{1/2},
\label{eq:orb1}
\end{equation}
and
\begin{equation}
P_\Phi= L=\left[{-J-(J^2-4IK)^{1/2}\over 2I}\right]^{1/2},
\label{eq:orb2}
\end{equation}
where
\begin{eqnarray*}
I&\equiv& {\tilde A^2r^4\over G^2}-{r^2\over G}(2\tilde A C
+B^2)+C^2, \\
J&\equiv& {(D+F)\over G}\Bigl[{2\tilde A^2r^2\over G}-2\tilde A C
-B^2\Bigr]+2\Delta (r^2\mu_o^2+Q)\Bigl[C-{\tilde A r^2\over G}
\Bigr], \\
K&\equiv& {\tilde A^2\over G^2}(D+F)^2-{2\tilde A \Delta\over G}
(r^2\mu_o^2+Q)(D+F)+\Delta^2(r^2\mu_o^2+Q)^2;
\end{eqnarray*}
and
\begin{eqnarray*}
\tilde A&\equiv& [(r^2+a^2)^2-a^2\Delta],\\
B&\equiv& 4Mar, \\
C&\equiv& \Delta-a^2, \\
D&\equiv& (3r^4-4Mr^3+a^2r^2)\mu_o^2, \\
F&\equiv& (r^2-a^2)Q,\\
G&\equiv& 3r^4+a^2r^2,
\end{eqnarray*}
for direct orbits of constant radius $r$, where 
$\Delta\equiv r^2-2Mr+a^2$.                                       

Upon applying equations~(\ref{eq:escape1}) and~(\ref{eq:escape2})
 to the Penrose pair production
(\gggg) electrons, for $E\equiv E_\mp$, $L\equiv L_\mp$, and 
$Q\equiv Q_\mp$, where $E_\mp$, $L_\mp$, $Q_\mp$ are the parameters
for a Penrose pair produced electron, we find that most satisfy the 
condition to have a turning point at the
iso-energy orbit $E_{\rm orb}$, with
$E_\mp=E_{\rm orb}$ and $L_\mp\ga L_{\rm orb}$ at  radii
$r_{\rm orb}\sim r_{\rm mb}$ (the last bound orbit for a
material particle,
deep within the ergosphere), before escaping to
infinity along vortical orbits about the polar axis,
 satisfying (Williams 1995)
\begin{equation}
0<{Q_\mp\over E_\mp^2}<{Q_{\rm orb}\over E_{\rm orb}^2},
\label{eq:escape3}
\end{equation}
or $Q_\mp<Q_{\rm orb}$, implying no turning point in $(P_\mp)_\Th$,
i.e., $(P_\mp)_\Th\nrightarrow 0$, yet $(P_\mp)_r\rightarrow 0$ at
$r_{\rm orb}$ according to satisfaction of equation~(\ref{eq:escape1}),
where $(P_\mp)_\Th$ is the polar coordinate angular momentum of a Penrose
produced electron. 

\section{Discussion}

That the Penrose produced $e^-e^+$ pairs escape to infinity along 
vortical trajectories without any appreciable interaction with 
stable accretion disk particles can be seen in the momentum spectra 
of Figures~1, 2 (for a supermassive KBH), and Figure~3 (for a 
``micro-massive'' KBH), as explained in the paragraphs below.  Note,
in gravitational units ($G=c=1$) such distribution spectral plots for the 
same initial energies are approximately the same irrespective of the 
mass of the black hole adopted to make the plot (see Williams 2002b).  
The Penrose pair 
production processes used to calculate Figures~1, 2, and~3 
are for an idealized configuration with the purpose to simply 
show the effect of the inertial frame dragging [produced 
by the gravitational field of the spinning black hole, i.e., its 
gravitomagnetic field (Williams 1999; 2002a)], and not an attempt to 
connect the calculations to detailed observations.   Nevertheless, the 
initial energies used for these Penrose processes are based on energies 
found to exist in relativistic accretion disk models (Novikov \& Thorne 
1973;  Eilek 1980; Eilek \& Kafatos 1983) for black hole masses 
$10^8 M_\odot$ and $30 M_\odot$, and the expected blueshift in orbital 
energies: increased by the factor $e^{-\nu}=\sqrt{-g^{tt}}\simeq 52$ at 
the photon orbit (Williams 1995), where $g^{tt}$ is the contravariant 
diagonal time component of the Kerr metric in Boyer-Lindquist (1967) 
coordinates; such initial energies yield final results consistent with 
general observed jet particle energies.  The initial incident photons, 
indicated by the subscript $\gamma 1$, are assumed to infall radially 
along the equatorial plane [$(P_{\gamma 1})_r<0$, 
$L_{\gamma 1}=Q_{\gamma 1}=0$), with energies $E_{\gamma 1}=0.03$~MeV 
for Figures~1 and 2; $E_{\gamma 1}=0.0035$~MeV for Figure~3.  These 
incident photons are allowed to collide with highly blueshifted target 
photons that orbit at the photon orbit $r=r_{\rm ph}=1.074M$, in 
nonequatorially confined trajectories, which repeatedly cross (i.e., 
pass through) the equatorial plane in ``spherical-like'' orbits that 
reach certain latitudinal angles (Wilkin 1972).  The collision is 
assumed to take place when the target photon passes through 
the equatorial plane.  The initial orbital parameters for the 
target photons, indicated by the subscript $\gamma 2$, are given by 
equations~(\ref{eq:orb1}) and~(\ref{eq:orb2}) for a particular 
$Q_{\gamma 2}$.  Specifically, the target photon orbits can be populated 
by prior Penrose Compton scattering (Williams 1995; 2002b), and probably 
by proton-proton scatterings of neutral pions that decay 
($\pi^0\lra\gamma\gamma$), occurring in ``hot'' ADAFs (Eilek \& Kafatos 
1983; Mahadevan, Narayan, \&  Krolik 1997), assuming some of the 
$\pi^0-$decay photons are created with appropriate $Q$ values and 
energies to become bound at the photon orbit (Williams 1995, 2002a).  
Now, in a realistic situation the incident and target particles are not 
constrained to this idealized configuration, but this configuration,
including the range of initial energies used (see also captions of
Figs.~1, 2, and~3), is most efficient and allows maximum energy-momentum 
extraction (Williams 1995).  

Figures~1$a$, 2$a$, and~3$a$ show the azimuthal momenta $L_\mp$ of the 
Penrose pair production
 (\gggg) versus
the energy $E_\mp$ of the escaping electron pairs after  2000 events [each
point (i.e., tiny dot) represents an escaping particle resulting from 
the events]. 
In general, since the energy $E$ as measured by an observer at
infinity is linearly proportional to the azimuthal angular momentum $L$
in the transformation laws, from the local nonrotating frame to the
observer's frame at infinity [compare eqs. (2.7c) and (2.7d) or (3.89a)
and (3.89d) of Williams 1995], and since there is a linear relationship
between $E$ and $L$ in the Boyer-Lindquist coordinate frame, 
as can be seen in 
Figures 1$b$, 2$b$, 
and 3$b$ (compare also equation~[\ref{eq:orb1}]), which is apparently
transferred to the scattered particles, the scatter plots (Figs.~1$a$, 
2$a$, 3$a$) representing 
the approximately 2000 escaping electrons per plot appear  linear.

Figures~1$b$, 2$b$,
and~3$b$ show the  azimuthal coordinate momentum versus energy 
of the $e^-e^+$ pairs, 
superimposed on the azimuthal coordinate momentum versus energy, 
orbital parameters, of an
electron orbit not confined to the equatorial plane 
(Wilkins 1972), at 
$r_{\rm mb}$ and $r_{\rm ms}$, the radii of marginally
bound and marginally stable orbits, respectively.  The 
conserved orbital
parameters $E$ and $L$ for the nonequatorially confined orbits
(for massless and material particles), which 
reduce to the forms of Bardeen, Press, Teukolsky (1972) 
for $Q=0$, as measured by
an observer at infinity, are given by
equations~(\ref{eq:orb1}) and~(\ref{eq:orb2}). 
Notice on Figures~1$b$, 2$b$, and~3$b$ 
that for $E_\mp=E_e\equiv E_{\rm orb}$ at 
$r_{\rm mb}$, $r_{\rm ms}\equiv r_{\rm orb}$ most 
of the escaping $e^-e^+$ pairs have $L_\mp>L_e$, indicating
 turning points 
inside the inner radius of the stable 
accretion disk ($\sim r_{\rm ms}$); and some, particularly the 
higher energy pairs, have turning points inside $r_{\rm mb}$.   
Moreover, compare Figures~1$b$, 2$b$, and~3$b$ over the total 
energy range to see that the curves for $r_{\rm ms}$ and 
$r_{\rm mb}$ do not merge into one curve at lower energies; they 
only appear to on Figures~1$b$ and~2$b$.  This, as well as 
the satisfaction of the turning point condition stated above 
(according to
equation~[\ref{eq:escape1}])
at $r\la$ $r_{\rm mb}$ and/or $\la$ $r_{\rm ms}$, can clearly be
seen when comparing Figures~4 and~5, which show specific intervals 
for the cases displayed in Figures~1$b$ and~2$b$.  

Notice also that,  the 
display in Figures~1$c$,
2$c$, and~3$c$, showing  
$\vert(P_\mp)_\Theta\vert<\vert(P_e)_\Theta\vert$ for
$E_\mp=E_e\equiv E_{\rm orb}$ at $r_{\rm mb}$, 
$r_{\rm ms}\equiv r_{\rm orb}$ for most 
of the escaping $e^-e^+$ pairs\footnote{The absolute
value brace are included to emphasize that we are comparing the
magnitude of the polar coordinate angular momentum.}, means that 
these electrons escape to
infinity along vortical trajectories,
 with helical angles of escape,
measured relative to the equatorial plane,
$\delta_\mp=\vert 90^\circ -\Theta_\mp\vert\sim 60^\circ$ to $0^\circ.5$, 
$\delta_\mp\sim 9^\circ$ to $0^\circ.5$, and 
$\delta_\mp\sim 6^\circ$ to $0^\circ.5$, for the electrons 
displayed in Figures~1, 2, and~3,
respectively (see Fig.~6; see also Williams 2002b, 
2002a).

Finally, notice that in Figure~1$c$, some of 
the $e^-e^+$ pairs with $E_\mp \la 2.5$~MeV do not escape and can
possibly  become bound, i.e., populate the orbits,
 at $r_{\rm mb}$ or $r_{\rm ms}$ for ``later''
Penrose Compton scattering events, involving these nonequatorially 
confined electrons as targets
(Williams 1995, 2002b). 
This shows that these Penrose processes and the accretion disk 
assumed (Williams 1995) can compile a self-consistent model 
producing the three 
general spectral high energy regimes observed in
 all AGNs, more or less (e.g., compare Figure~8 of Williams 2003).  
That is, for a 
particular self-consistent case,  somewhat {\it similar} to that of 
Figure~1, from prior Penrose 
Compton 
scattering of initial photons ($0.03$~MeV) by equatorially
confined electron targets at $ r_{\rm mb}$,
the final inward scattered photons 
$\sim 330$~keV, 
with turning
points after being blueshifted by a factor of $e^{-\nu}\simeq 52$ 
(Williams 2002a), 
can populate the photon orbit for
Penrose pair production (\gggg), which in turn yields electrons
that can populate the nonequatorially confined target orbits: with 
energies up to $\sim 3$~MeV, similar to that seen in Figure~1$c$,
for subsequent Penrose 
Compton scattering.  In this self-consistent case, however,
$E_\mp$ can extend
up to $\sim 17$~MeV (compare Fig.~1).
Note,  the population of the
target particle orbits and accretion disk properties are discussed in 
details in Williams (1995, 2002a, 2002b, 2003).               
  
The helical angles of escape stated above (compare Fig.~6), 
and given analytically
by 
\begin{equation}
\dd=\arccos{\left[ {-T+\sqrt{T^2-4SU}\over 2S} \right] }^{1/2},
\label{eq:helical1}
\end{equation}
where
\begin{eqnarray*}
S&\equiv &a^2(\mu_o^2-E^2), \\
T&\equiv &Q +a^2(E^2-\mu_o^2)+L^2,\\
U&\equiv &-L^2
\label{eq:helical2}
\end{eqnarray*}
[as derived from 
the ``Carter constant'' of motion by letting $P_\Theta\longrightarrow 0$
(see equation~[\ref{eq:carter}]; see also Piran \& Shaham 1977), 
where $\Theta_\mp=90^\circ\pm \dd_\mp$ for escaping below ``$+$''
or above ``$-$'' the equatorial plane],
suggest that the particles escape
to large distances 
along coil-like geodesics conglomerated in the form of  collimated
swirling
current plasma polar jets, with energy-momentum of the KBH transferred
from the highly blueshifted
frame dragged scattering events. 
Now, a test particle moves on a geodesic in a gravitational field
if it is not acted on by some external force, but if the parameters such
as energy and angular momentum vary with time, then
forces are present that can destroy the geodesic character of the motion.
But the geodesic character of a trajectory, however, can be approximately
saved if the time scale of a significant variation of the physical
parameters, say $\tau_{var}$, is longer than the dynamical time, $\tau_{dyn}$,
associated with a geodesic trajectory (de Felice \& Carlotto 1997).
This means that, if the orbital parameters vary slightly so as to
preserve the initial $\Th$ angle (i.e., the emission angle of the particle
above or below the equatorial plane), then under suitable conditions, say
in a suitable electromagnetic field, we
obtain a net axial collimation.   

Calculations by de Felice \& Carlotto (1997); Karas \& Dov\v{c}iak
(1997) suggest that
the average
dynamical time associated with the vortical geodesics is
sufficiently short to give confidence that the average time
of a significant variation of the physical parameters,
due to dissipative or accelerating processes, is long enough
to ensure the geodesicity condition on outgoing vortical
orbits. This then can possibly  permit the maintenance of the intrinsic
collimation of the Penrose scattered particles by the KBH, particularly
because of the sturdiness or stiffness (de Felice \& Carlotto 1997) of
the coil-like structure of the geodesics.
Moreover, their calculations also show that for a Lorentz force
acting on a charge particle near the event horizon, for a $10^8 M_\odot$
KBH, inertia and gravity dominate over electromagnetic disturbance.     

For completeness, the overall difference  in the distributions
of the polar angles of escape, displayed in Figure~6,
is a general relativistic effect due to the gravitomagnetic (GM) 
force field acting on the
momentum of individual particles, resulting in an ``alteration''
of the incoming and scattering angles.  The resultant
force in the polar direction, as conveyed in the spectra of Figure~6, 
depends largely
on the counterbalance between the dynamics of the nonequatorially confined
target photons' orbits (with latitudinal angles $\la 0^\circ.5$
about the equatorial plane) and the GM force. 
Note, an important feature of the GM
field (caused by the angular momentum of the rotating black
hole)
is the production of symmetrical and asymmetrical polar jets: 
appearing to break the expected reflection symmetry of the Kerr metric,
above and below the equatorial plane (Williams 2002a).  The general
shapes of the spectra are not the same because the 
energy-momenta of
the initial and escaping
particles, in the three cases (Figs.~6$a-6c$) are different; therefore,
the GM force (${\bf F}_{_{\rm GM}}={\bf\tensor \cdot p}$),
acting proportional to the momentum ($\bf p$) of a particle, 
will be different,
where $\bf\tensor$ is the GM tensor (Thorne,
Price, \& Macdonald 1986).  The GM force behaves in some degree like
a ``Coriolis
force,'' acting on the momentum of particles in the 
local inertial frame that has been dragged into rotation.
This can be somewhat understood by looking at the GM force component
in the polar direction:
\begin{equation}
(F_{_{\rm GM}})_\Theta\propto (H^r P_\Phi
-\tilde H^\Phi P_r);
\label{eq:GM}
\end{equation}
with
\begin{eqnarray}
&H^r\leq 0 ~\,{\rm for}~\,
\Theta\leq 90^\circ, \nonumber \\
&H^r\geq 0 \,~{\rm for}~\,\Theta\geq 90^\circ,
  \nonumber \\
&\tilde H^\Phi>0~\,{\rm for}\,~ \Theta\leq 90^\circ~\,{\rm and}~\,
 \Theta>90^\circ,
\label{eq:GM1}
\end{eqnarray}     
as defined by the
 components of the
GM tensor for $r\sim r_+$, where it appears that the spacetime frame dragging
is so severe that the GM field lines are 
distorted into the direction of rotation,
producing the nonzero $\tilde H^\Phi$ term
 (Williams 2002a).  Notice in equation~(\ref{eq:GM}), using
equation~(\ref{eq:GM1}),
that the first term maintains equatorial reflection symmetry, while the 
second term introduces asymmetry; this second term is expected 
to exist effectively only near the event horizon ($r_+$), i.e., inside the 
ergosphere: 
the region of importance for
the Penrose mechanism to operate.
A general analysis of equations~(\ref{eq:GM}) and~(\ref{eq:GM1})  
reveals the following conditions:
\begin{itemize}
\item[1.] At $\Theta=90^\circ$, $P_r<0$ or $P_r>0$, the force on the
particle will be $(F_{_{\rm GM}})_\Theta>0$ or $(F_{_{\rm GM}})_\Theta<0$,
respectively.
\item[2.] At $\Theta>90^\circ$, $P_r>0$, $P_\Phi>0$, the force on the
particle will be $(F_{_{\rm GM}})_\Theta>0$ for $H^r P_\Phi>
\tilde H^\Phi P_r$, and $(F_{_{\rm GM}})_\Theta<0$ for $H^r P_\Phi<
\tilde H^\Phi P_r$.
\item[3.] At $\Theta<90^\circ$, $P_r>0$, $P_\Phi>0$, the force on the
particle will be $(F_{_{\rm GM}})_\Theta<0$. 
\end{itemize}

We now use the three conditions above to explain the 
escaping particle
distribution in the individual 
spectra of Figure~6.  The momentum components ($P_\Phi$, $P_r$) 
and the polar angle ($\Theta$) are
to be compared with that of the initial and escaping particles.
In the case of the lowest energy (Fig.~6$c$; see also Fig.~3)
target ($E_{\gamma 2}\simeq 3.387$~MeV) and incoming 
($E_{\gamma 1}=0.0035$~MeV)
photons, the initial asymmetry, favoring the positive $\unitt$ direction,
is established by the GM field acting on the radially infalling photons
according to condition~1.  The outward radial momenta of the escaping
electrons with  $90^\circ\la\Theta_\mp\la 90^\circ$, according to 
conditions~1$-$3, are not large enough to overcome the force produced by 
$(F_{_{\rm GM}})_\Theta>0$; therefore, the asymmetry remains.  
For the lowest energy escaping electrons, the GM force dominates over the
dynamics of the target photons, resulting in helical angles of escape
$\delta_\mp> 0^\circ.5$ for most of the escaping electrons.  
Also, the $H^r$
component has a larger absolute value for a particle with a larger
$\delta_\mp$ (Williams 2002a), thus contributing to 
the distribution
being forced away from the equatorial plane according to conditions~2 and~3
(compare eqs.~[\ref{eq:GM}] and~[\ref{eq:GM1}]).
Finally, in this particular case, as the energy of the escaping 
electrons increases, 
the GM force
dominates less and less, while the orbital dynamics of the target
photons begin to dominate, producing  smaller $\delta_\mp$ values, 
consistent
with the latitudinal angles of the target photons
(which are $\la 0^\circ.5$). 

In the case of the somewhat higher energy 
(Fig.~6$a$; see also Fig.~1)
target ($E_{\gamma 2}\simeq 13.54$~MeV) and incoming
($E_{\gamma 1}=0.03$~MeV)
photons, the initial asymmetry, favoring the positive $\unitt$ direction,
caused by the GM field acting on the radially infalling photons
according to condition~1 appears to be approximately
balanced by $(F_{_{\rm GM}})_\Theta<0$ according to conditions~2 and~3, 
except at the very highest energy of the escaping electron
pairs, where $(F_{_{\rm GM}})_\Theta>0$ dominates according to condition~2,
producing asymmetry.  
For most of the  escaping electrons, the GM force dominates over the
dynamics of the target photons, resulting in 
helical angles of escape
as high as $\delta_\mp\sim 60^\circ$ according to conditions~2 and~3.  
Nevertheless, as the energy 
of the escaping electrons increase, the orbital dynamics of the target
photons begin to dominate, producing  smaller $\delta_\mp$ values.
                              
In the case of the highest energy
(Fig.~6$b$; see also Fig.~2)
target ($E_{\gamma 2}\simeq 2146$~MeV) and incoming
($E_{\gamma 1}=0.03$~MeV)
photons, the asymmetry, favoring the positive $\unitt$ direction,
caused by the GM field acting on the the radially 
infalling photons
according to condition~1 and the high energy escaping electrons 
according
to condition~2, where $(F_{_{\rm GM}})_\Theta>0$ dominates
 for $H^r P_\Phi>
\tilde H^\Phi P_r$ (recall that $P_\Phi$ increases with increasing $E$), 
appears to dominate over $(F_{_{\rm GM}})_\Theta<0$
in conditions~1$-$3. 
Also,  except for the relatively very lowest 
energy of the 
escaping pairs, the orbital dynamics of the target photons
appear to dominate over the GM force, as the escaping energy increases, 
producing  predominantly 
electrons with $\delta_\mp\la 1^\circ$.

Note, in these Penrose processes, the GM force is inherently tied 
to the overall scattering
process through the Kerr metric.  That is, just as we do not have to 
calculate separately the gravitational
force on the scattered particles due to the mass $M$ of the KBH,
we do not have to calculate separately
the GM force on the scattered particles due to the angular momentum
$J$ of the KBH.  
The GM  force, however, is
given here separately only to show how it relates to the final
four-momenta of the scattered particles, in an effort to explain the
major difference of spectral distributions displayed in Figure~6.
                                                                                
\section{Conclusions}

In this paper, I have shown that the Penrose pair production
 (\gggg) electrons,
of processes occurring at the photon orbit $r_{\rm ph}$,
escape to infinity along vortical trajectories about the polar 
axis, without any  appreciable interaction with the 
stable accretion disk particles (which are located at $r\ga r_{\rm ms}$). 
  This is also expected to be true for the Penrose Compton 
scattered escaping
photons that have inward directed radial momenta
with turning points at $r\sim r_{\rm ph}$, escaping along 
vortical orbits concentric to the polar axis (Williams 2001,
2002b).  On the other hand, since the Penrose Compton 
scattering processes
occur at radii $r_{\rm mb}\la r\la r_{\rm ms}$ the scattered
photons with positive radial momenta probably have
appreciable interaction with the inner region of the
accretion disk, 
as suggested by the broad Fe K$\alpha$
emission line at $\sim 6$~keV observed in the bright Seyfert~1
galaxy MCG---6-30-15 (Wilms et al.~2001); see the qualitative 
description in Williams (2002b). 
Details concerning MCG---6-30-15 will be investigated in
a future paper.

Overall, these vortical trajectories of escaping particles
suggest that the KBH is 
responsible for 
the precollimation (de Felice \& Curir 1992;
de Felice \& Carlotto 1997;  de Felice \& Zanotti 2000;
see also Bi\v{c}\'{a}k, Semer\'{a}k, \& Hadrava 1993;
Karas \& Dov\v{c}iak 1997) of the observed jets of relativistic
particles, 
emanating from the cores of objects  powered
by black holes (compare Junor, Biretta, \& Livio 1999).  
Note, the
coil-like collimation of the vortical orbits, presented here,
and those of the Penrose Compton scattering processes,
are investigated in further details 
elsewhere.  Further investigations of this 
precollimation will include the self-induced dynamo magnetic
field associated with the vortical orbiting escaping Penrose
produced (\gggg) pairs; such a field could be an important contribution 
to the observed synchrotron emission as well as  in maintaining 
collimation.  Recent polarization measures (e.g., see Homan 2004 and 
references therein) appear to be consistent with the dynamics and 
kinematics of these vortical polar trajectories of escaping electrons.
The degree to which the above statement applies consistently, with the 
interpretation that the
observed transverse rotation-measure gradients
across the jets (Gabuzda \& Murray 2003) are due to an intrinsic helical
magnetic field structure associated with the accretion disk, must be 
looked at in detail.  The best-case scenario would be for the magnetic 
field of the accretion disk to assist in further collimating and 
accelerating
the Penrose produced relativistic jet particles as they escape from the 
black hole out to observed distances.

In comparison with some MHD energy extraction models, of the Blandford-Znajek
(1977) and Blandford-Payne (1982) type, we find the following.
Firstly, because of the proximity of the Penrose pair production 
(\gggg) processes
(Williams 1995) to the event horizon, i.e., occurring at the photon orbit
$r_{\rm ph}$ inside the marginally bound orbit $r_{\rm mb}$ (the last radius
for any bound material particle before it falls directly into the black hole),
the accretion disk magnetic field, proposed to be frozen to the plasma
particles (e.g., Koide et al. 2002;  Koide  2003; Meier,
Koide, \& Uchida 2001), is expected to have a negligible effect on these pair
production processes, even without considering that the existences of such
magnetic fields or ``flux tubes'' near the event horizon are inconsistent with
general relativistic findings by Bi\v{c}\'{a}k (2000) and Bi\v{c}\'{a}k \&
Ledvinka (2000).  These findings reveal that such fields, if aligned with the
rotation axis, will be expelled from the horizon or redshifted away, for a
rapidly rotating black hole (i.e., a near extreme KBH).  Now based on these 
findings, it is probably safe to say that the disk magnetic field may have
little, if any, effect on the Penrose Compton scattering processes
(Williams 1995), particularly those occurring at $\sim r_{\rm mb}$, the radii 
first to be populated by the target electrons (Williams 2002b) and where 
the most energy would be 
extracted for a specific $Q_e$ value [compare the bound electron orbits
displayed in Figs.~1$c$ and~3$c$; see also Fig.~1(b) of Williams 1995].
The radius of marginally bound orbit $r_{\rm mb}$ is closer to the 
event horizon ($\Delta r\equiv r_{\rm mb}-r_+\simeq 0.026M$) 
than it is to $r_{\rm ms}$ 
($\Delta r\equiv r_{\rm mb}-r_{\rm ms}\simeq 0.111M$), 
where $r_+\simeq 1.063M$ for $a=0.998M$ (Bardeen et al. 1972).

Secondly, the main advantage of
the Penrose-Williams processes, over such Blandford-Znajek (1977) and
Blandford-Payne (1982) type models, is that the Penrose-Williams gravitational
energy-momentum extraction processes occur independently of a magnetic field,
while producing highly relativistic particles that escape to infinity along
intrinsically collimated vortical trajectories in the form of symmetrical or
asymmetrical jets, because of the inertial frame dragging.  Another feature
of these processes is that
most of the energy extracted is gravitational binding energy with only a small
fraction of the scattered particles having final negative energies (up to
$\sim 30$\% of the Penrose Compton scattered photons and up to $\sim 10$\% of
the Penrose pair produced electrons, occurring only in the lowest
energy-momentum scattering processes), which means that the angular momentum
of the black hole will not decrease significantly in these processes.  The 
lifetime of these processes is expected to be indefinite as long as there 
is matter to be accreted, since positive energy-momentum particles are
also scattered into the KBH. The resulting escaping plasma jets are expected 
to generate a self-induced dynamo magnetic field, as stated above, in the form
of a polar solenoid-like field, which could possibly magnetically confine and
further assist in acceleration, collimation processes, as well as producing
observed synchrotron radiation.  Importantly, as soon as the inner region
unstable disk particles reach a temperature $T\ga 17$~keV---the energy needed
to populate the target particle orbits for Penrose Compton scattering,
assuming, of course, the probable existence of turning points
(compare eq.~[\ref{eq:escape1}] and discussion in \S~\ref{sec:orbits}), this
Penrose process can ``turn on'' (Williams 2002b), irrespective of the mass
of the rotating black hole.  Moreover, the general observed high energy 
luminosity spectra of quasars and microquasars can be reproduced by these
processes (see Williams 2002b, 2003).  The above features are general 
characteristics of observed black hole sources.  None of the 3-D MHD energy 
extraction jet models at the present (e.g., Koide et al. 2000; Meier et al. 
2001; Koide et al. 2002; Koide 2003, 2004) exhibit such characteristic 
features in such details as 
does the Penrose-Williams mechanism.  Nevertheless, 3-D MHD models can
possible achieve the necessary powers (Blandford \& Znajek 1977;
Meier et al. 2001), but, in some cases, only after assuming an 
unrealistically large strength magnetic field.                                                   
 
Lastly, it appears that the modern-day Blandford-Znajek (1977) and 
Blandford-Payne
(1982) type models are essentially faced with the age-old problem of
converting from electromagnetic energy to particle energy and thus the
inability of generating the highly relativistic particles needed to be
consistent with observations.  The problems associated with such models
in the direct extraction of energy near the event horizon suggest that these
models may be important in the weak gravitational field limit, serving perhaps
the same purpose they do in the jets of protostars, i.e., appearing to have a
dominant role on a large scale at distances outside the strong effects of
general relativity.  The recent model of Koide (2004) and its application
to $\gamma$-ray bursts, assuming a large scale, superstrong radial magnetic
field ($\sim 10^{15}$~G) down to the event horizon of a rapidly rotating
KBH ($a=0.99995$), produces a mildly relativistic outflow.  Not only,
at least at the present, does this model not collimate, but it may not be
consistent with the general relativistic determination that any ``radiatable''
multipole field gets radiated away completely as a star collapses to a KBH
(de la Cruz, Chase, \& Israel 1970; Price 1972), which includes the magnetic
field, leaving only nonzero monopole parameters $M$, $J$, and $q$, where
$J$ ($=Ma$) is the black-hole angular momentum, and $q$ is the electric charge,
hence the ``no-hair theorem'' (Misner, Thorne, \& Wheeler 1973; Carter 1973).
Also, the problem remains in the model of Koide (2004) as to how such a
superstrong field is created in the popular (or generally accepted) model of
stellar evolution.
                                                                                
In any case, the Penrose-Williams processes described in this paper (see
also Williams  2002a, 2002b, 2003) may be related to the beamed energy
output observed in $\gamma$-ray bursts.  It is suspected that these
processes, which could ``quickly'' turn on and off under suitable conditions,
might be important.  Particularly, these processes might prove to be
invaluable in explaining the jets in so-called collapsars, i.e, black hole
formation in massive stars with rotation (MacFadyen, Woosley, \& Heger 2001),
without the needed of a superstrong magnetic field.  Equally, considering
an ``inactive'' and/or perhaps ``isolated'' rotating  black hole, if its
tidal forces encounter and destroy an object of sufficient density, then as
the debris is infalling, these Penrose processes could produce characteristics
associated with some $\gamma$-ray bursts: an energetic short-lived burst,
collimation, synchrotron emission, as well as the afterglows in the X-ray
(Costa et al. 1997), optical (van Paradijs et al. 1997), and radio 
(Frail et al.
1997) regimes.  That is, $\gamma$-ray and X-ray jets can  be
produced possibly from Penrose Compton scattering, and radio to optical 
synchrotron jet emission from subsequent Penrose pair production (\gggg) 
electrons, interacting with the expected intrinsically induced magnetic 
field (mentioned above).  
Details of the application of these Penrose processes to $\gamma$-ray bursts 
await future investigations.
 
Note, in general, evaluation in a full-scale relativistic MHD regime is 
needed to 
follow the evolution
of the trajectories of the escaping Penrose particles and their interactions
with the intrinsically induced magnetic field and that of
the surrounding accretion disk.  This indeed will be a challenge. Yet, with
existing MHD 3-D simulations, and since we have analytical expressions for
the trajectories of the escaping particles (see Williams 2002a), such a task
could be accomplished.
 
Finally, for completeness, even though the Penrose processes present 
here are quite efficient (Williams 1995) without a magnetic field, the 
presence of
a disk magnetic field, however, inside the ergosphere, might increase the
efficiency (see Wagh \& Dadhich 1989 and references therein).  The effects
of such a magnetic field, if sufficiently small, might be represented by a
random motion superimposed on the orbital velocities of the charged particles,
say the target electrons in Penrose Compton scattering (see e.g. Williams 
1995).

\acknowledgments

I first thank God for His thoughts 
 and for making this research possible.  
Next, I thank 
Dr.~Fernando de Felice and Dr.~Henry Kandrup for their helpful comments
and discussions.  Also, I thank Dr.~Roger Penrose for his continual
encouragement.  I am grateful to Dr. Ji\v{r}\'{i} Bi\v{c}\'{a}k for 
helpful comments. This work was supported 
in part by a grant from NSF and AAS Small Research Grant.


\clearpage


\begin{figure} 
\epsscale{.75}
\plotone{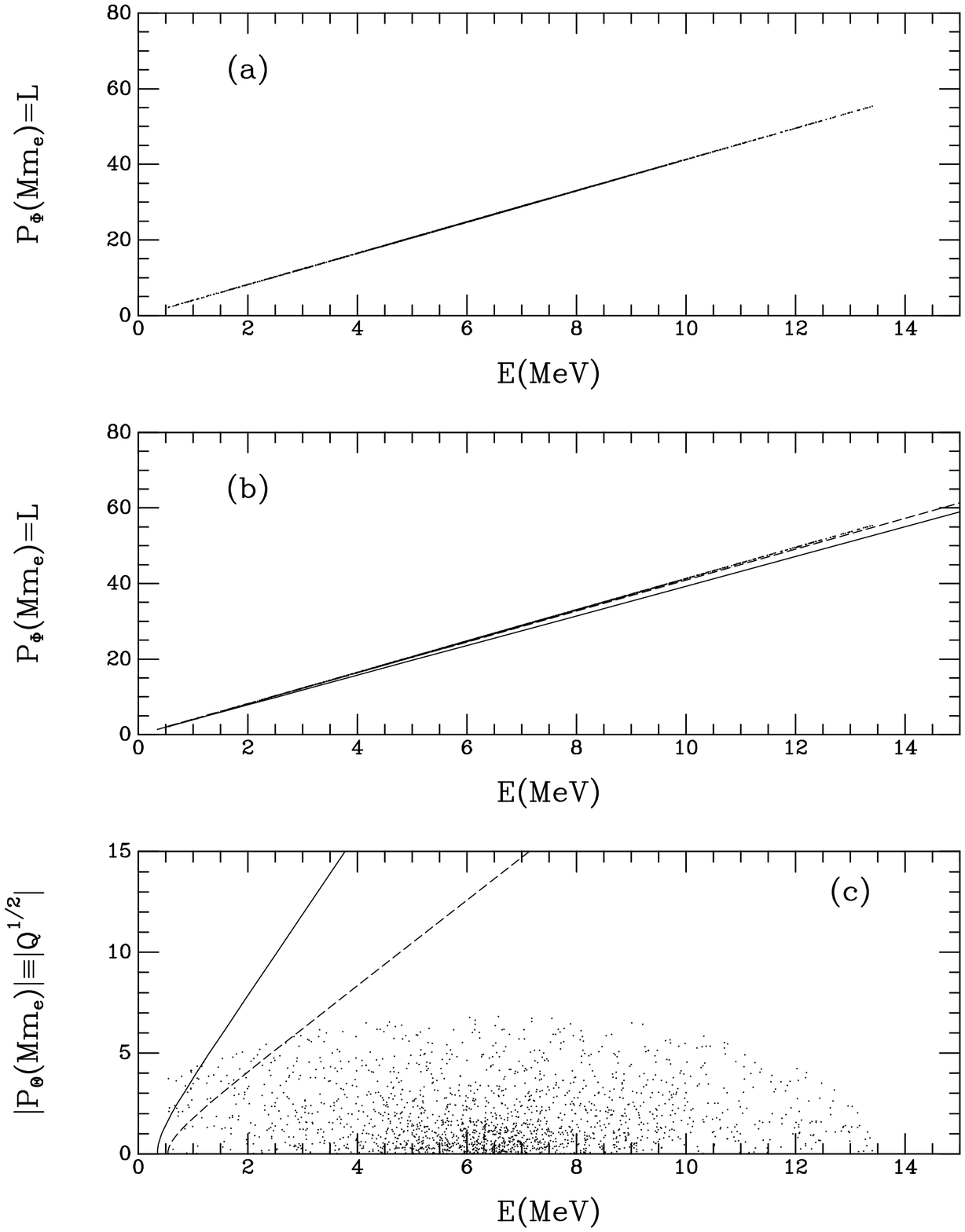}
\caption{Magnitude of the azimuthal ($P_\Phi=L$) and polar 
($P_\Theta$)
coordinate momenta for escaping vortical trajectories of electrons from  
Penrose pair production
(\gggg) at $r_{\rm ph}=1.074M$, and for the bound nonequatorially
confined electron particle orbits that cross the equatorial plane 
($P_\Theta\equiv Q^{1/2}$): $(P_{e})_\Phi$ vs.~$E_e$ and $(P_{e})_\Theta$ 
vs.~$E_e$, of electron orbits, at
$r_{\rm mb}\simeq 1.089M$ (dashed curve) and  at
$r_{\rm ms}\simeq 1.2M$ (solid curve). (a) Scatter plot displaying
$L_\mp$ vs. $E_\mp$ of $e^-e^+$ pairs after 2000 events (each point
represents an escaping electron).  The case shown has initial parameters:
$E_{\gamma 1}=0.03$~MeV, the infalling photon energy;
$E_{\gamma 2}\simeq 13.54$~MeV, the target photon orbital energy;
$L_{\gamma 2}\simeq 55.6 Mm_e$, corresponding azimuthal coordinate momentum;
$Q_{\gamma 2}^{1/2}=\pm 0.393\,Mm_e$, corresponding polar 
coordinate
momentum $(P_{\gamma 2})_\Theta$; and $M=10^8 M_\odot$.
(b) $L_\mp$ vs.~$E_\mp$ superimposed on the orbital parameters
$L_e$ vs.~$E_e$; see also Fig.~4. (c) $(P_\mp)_\Theta$ vs. $E_\mp$ 
superimposed on 
the orbital parameters
$(P_e)_\Theta$ vs.~$E_e$.
\label{fig1}}                        
\end{figure}
\clearpage
 
\begin{figure} 
\epsscale{.75}
\plotone{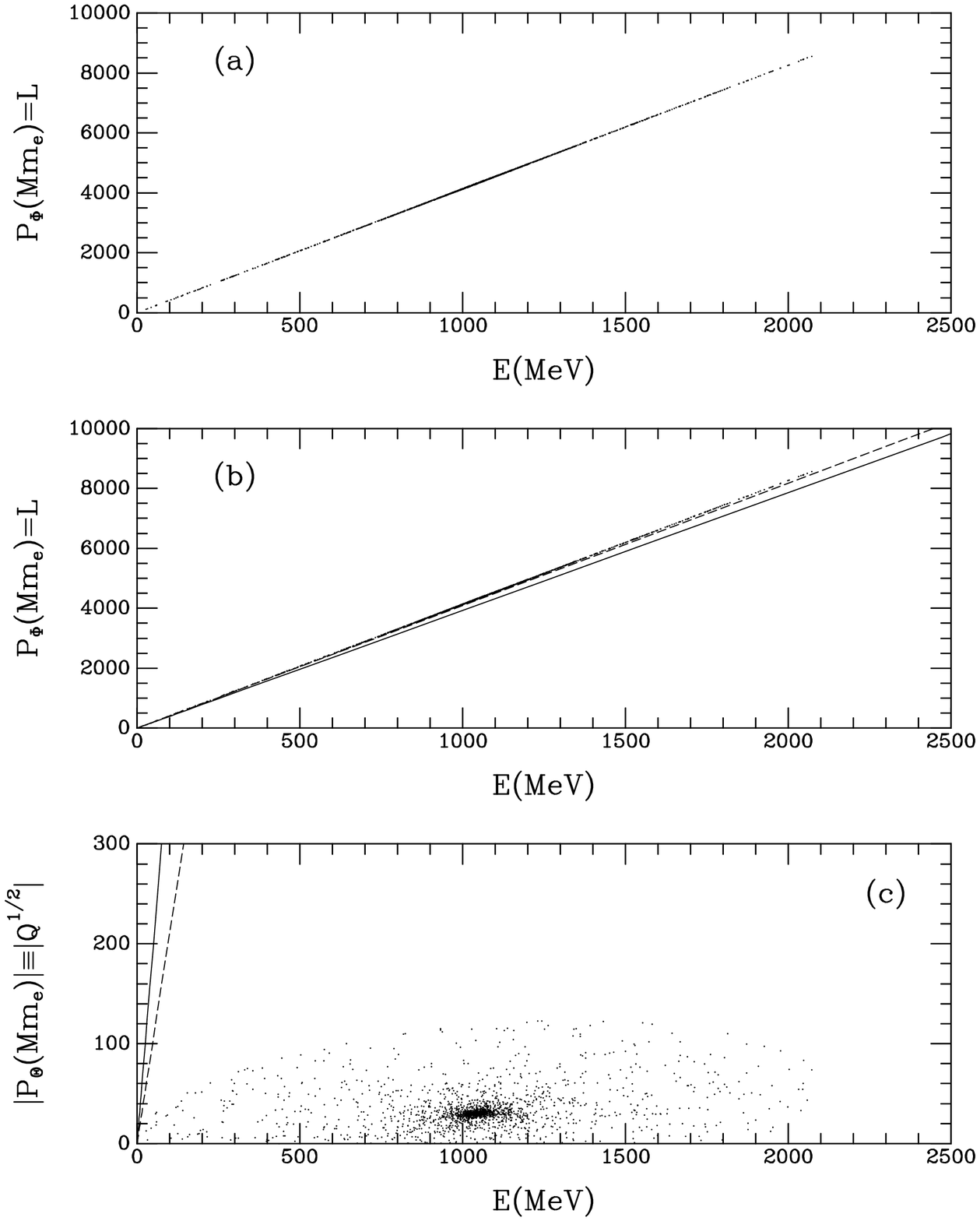}
\caption{Magnitude of the azimuthal ($P_\Phi=L$) and polar
($P_\Theta$)
coordinate momenta for escaping vortical trajectories of electrons from  
Penrose pair production
(\gggg) at $r_{\rm ph}=1.074M$, and for the bound nonequatorially
confined electron particle orbits that cross the equatorial plane
($P_\Theta\equiv Q^{1/2}$): $(P_{e})_\Phi$ vs.~$E_e$ and $(P_{e})_\Theta$
vs.~$E_e$, of electron orbits, at
$r_{\rm mb}\simeq 1.089M$ (dashed curve) and  at
$r_{\rm ms}\simeq 1.2M$ (solid curve). (a) Scatter plots displaying
$L_\mp$ vs. $E_\mp$ and $(P_\mp)_\Theta$ vs.~$E_\mp$
of $e^-e^+$ pairs after 2000 events (each point
represents an escaping electron).  The case shown has initial 
parameters:
$E_{\gamma 1}=0.03$~MeV, the infalling photon energy;
$E_{\gamma 2}\simeq 2146$~MeV, the target photon orbital energy;
$L_{\gamma 2}\simeq 8.81\times 10^3 Mm_e$, corresponding azimuthal 
coordinate momentum;
$Q_{\gamma 2}^{1/2}=\pm 62.28\,Mm_e$, corresponding polar
coordinate
momentum $(P_{\gamma 2})_\Theta$; and $M=10^8 M_\odot$.
(b) $L_\mp$ vs.~$E_\mp$ superimposed on the orbital parameters
$L_e$ vs.~$E_e$; see also Fig.~5. (c) $(P_\mp)_\Theta$ vs. $E_\mp$ 
superimposed on
the orbital parameters
$(P_e)_\Theta$ vs.~$E_e$.
\label{fig2}}
\end{figure}
\clearpage                                                                      
 
\begin{figure} 
\epsscale{.75}
\plotone{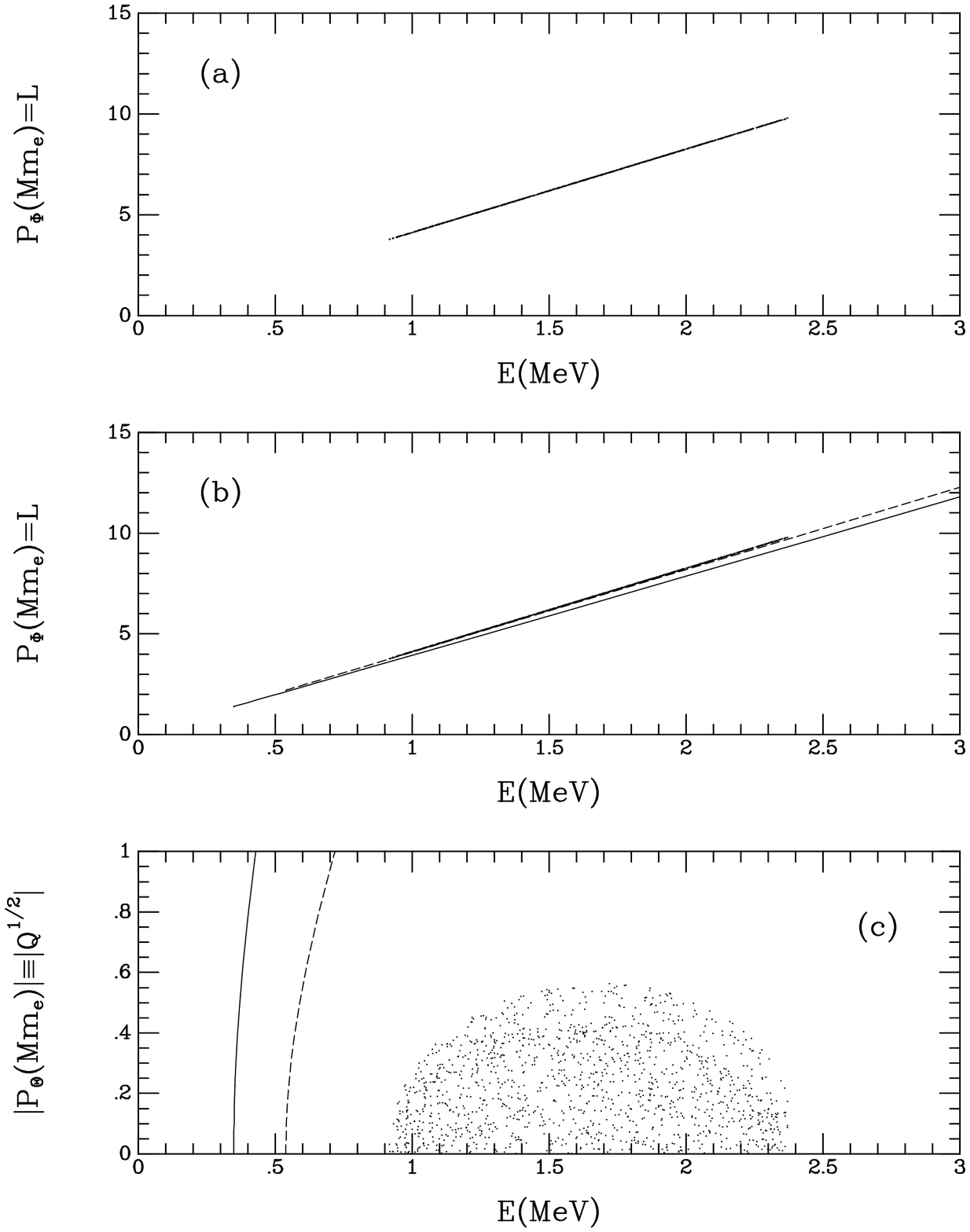}
\caption{Magnitude of the azimuthal ($P_\Phi=L$) and polar
($P_\Theta$)
coordinate momenta for escaping vortical trajectories of electrons from  
Penrose pair production
(\gggg) at $r_{\rm ph}=1.074M$, and for the bound nonequatorially
confined electron particle orbits that cross the equatorial plane
($P_\Theta\equiv Q^{1/2}$): $(P_{e})_\Phi$ vs.~$E_e$ and $(P_{e})_\Theta$
vs.~$E_e$, of electron orbits, at
$r_{\rm mb}\simeq 1.089M$ (dashed curve) and  at
$r_{\rm ms}\simeq 1.2M$ (solid curve). (a) Scatter plots displaying
$L_\mp$ vs. $E_\mp$ and $(P_\mp)_\Theta$ vs.~$E_\mp$ of $e^-e^+$ 
pairs after 
2000 events (each point represents an escaping electron).  
The case shown has initial parameters:
$E_{\gamma 1}=0.0035$~MeV, the infalling photon energy;
$E_{\gamma 2}\simeq 3.387$~MeV, the target photon orbital energy;
$L_{\gamma 2}\simeq 13.86 Mm_e$, corresponding azimuthal coordinate momentum;
$Q_{\gamma 2}^{1/2}= \pm 0.125\,Mm_e$, corresponding polar
coordinate
momentum $(P_{\gamma 2})_\Theta$; and $M=30 M_\odot$.
(b) $L_\mp$ vs.~$E_\mp$ superimposed on the orbital parameters
$L_e$ vs.~$E_e$. (c) $(P_\mp)_\Theta$ vs. $E_\mp$ superimposed on
the orbital parameters
$(P_e)_\Theta$ vs.~$E_e$.
\label{fig3}}
\end{figure}
\clearpage   

\begin{figure} 
\epsscale{.86}
\plotone{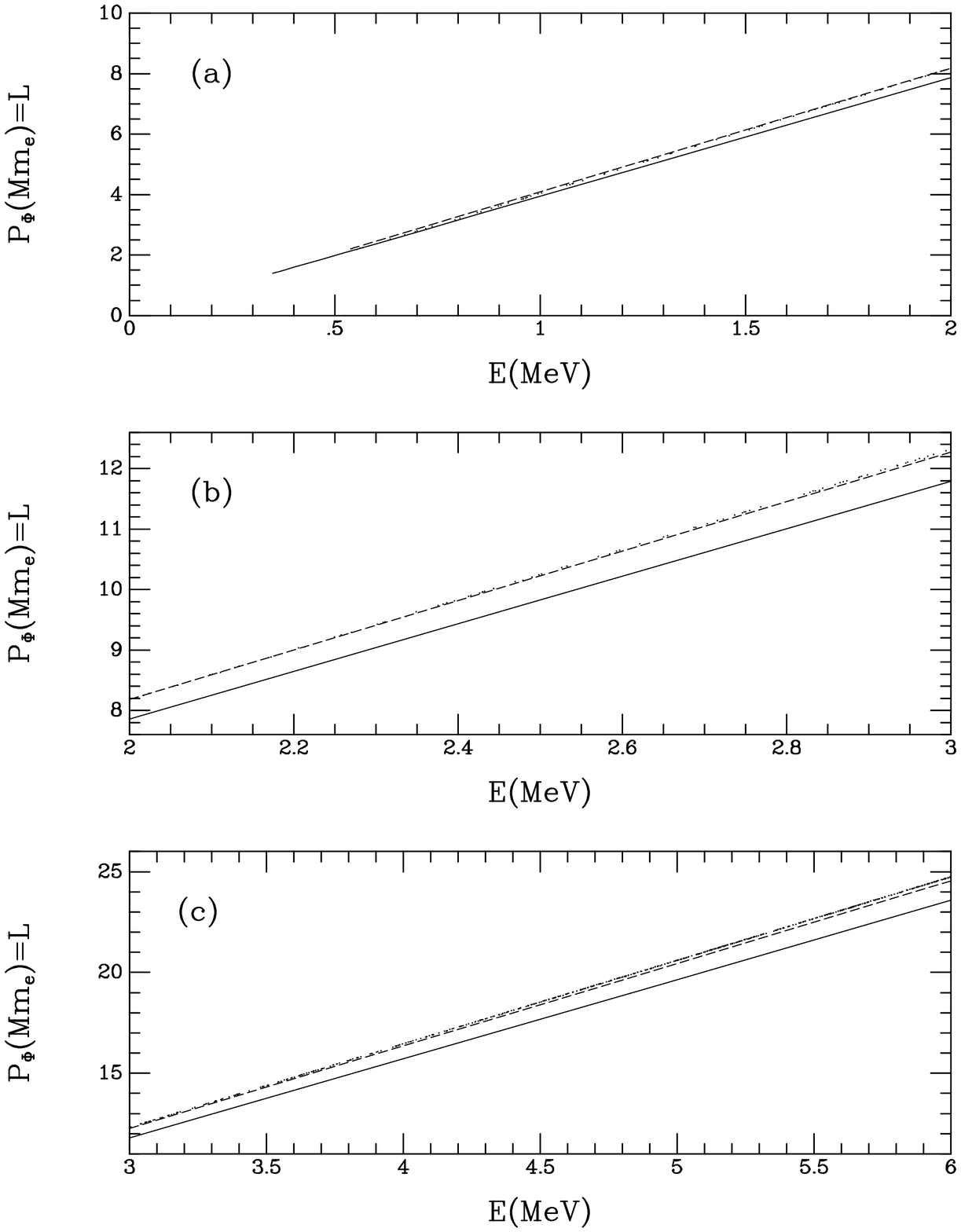}
\caption{Magnitude of the azimuthal ($P_\Phi=L$)
coordinate momentum for escaping vortical trajectories of 
electrons from
Penrose pair production
(\gggg) at $r_{\rm ph}=1.074M$ (scatter points), and for the bound 
nonequatorially
confined electron particle orbits that cross the equatorial plane: 
$(P_{e})_\Phi$ vs.~$E_e$, of electron orbits, at
$r_{\rm mb}\simeq 1.089M$ (dashed curve) and  at
$r_{\rm ms}\simeq 1.2M$ (solid curve). Panels (a), (b), and (c),
showing $L_\mp$ vs.~$E_\mp$, 
are specific intervals corresponding to Fig.~1$b$ (see text).
\label{fig4}}
\end{figure}
\clearpage

\begin{figure} 
\epsscale{.86}
\plotone{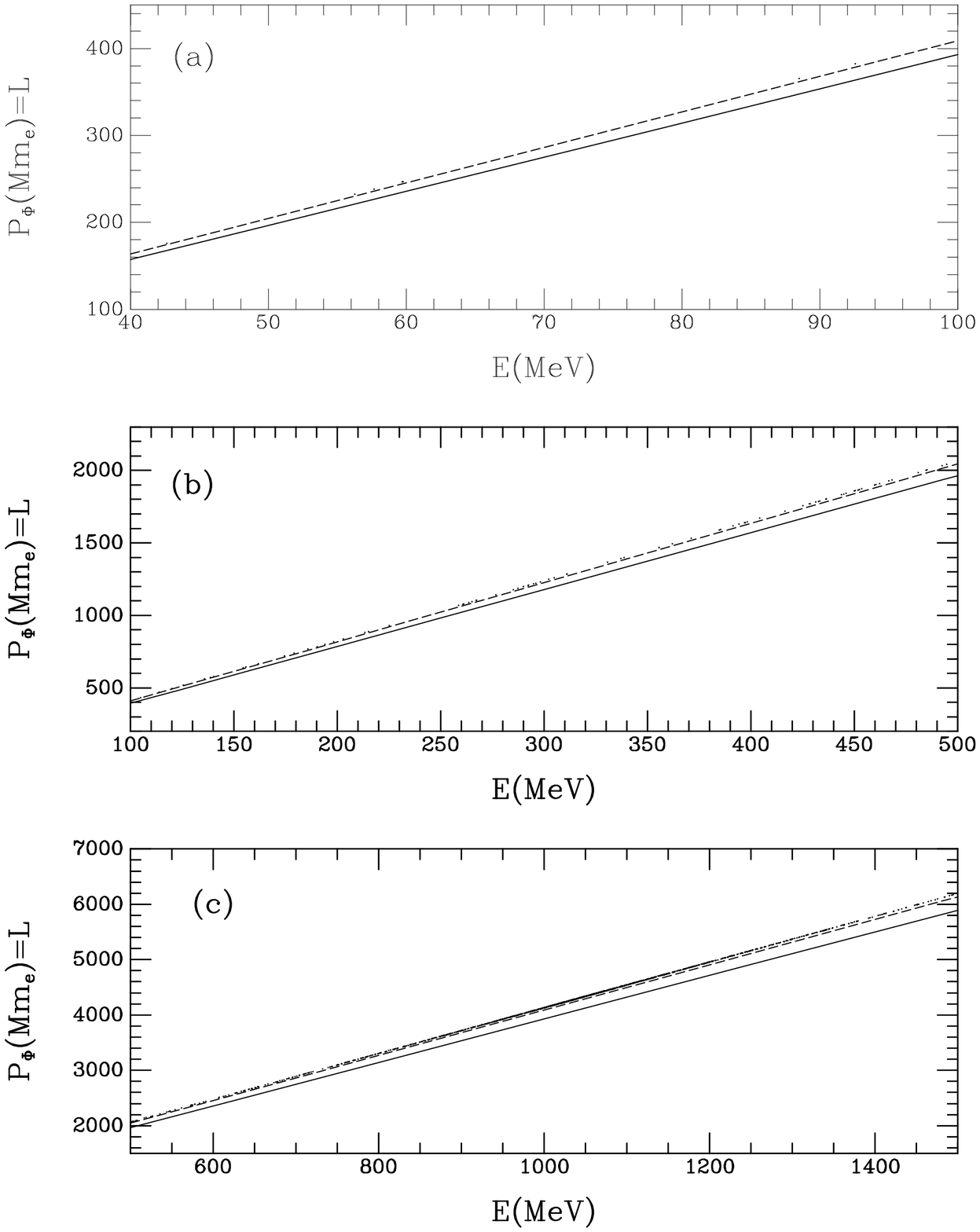}
\caption{Magnitude of the azimuthal ($P_\Phi=L$)
coordinate momentum for escaping vortical trajectories of 
electrons from
Penrose pair production
(\gggg) at $r_{\rm ph}=1.074M$ (scatter points), and for the bound 
nonequatorially
confined electron particle orbits that cross the equatorial plane:
$(P_{e})_\Phi$ vs.~$E_e$, of electron orbits, at
$r_{\rm mb}\simeq 1.089M$ (dashed curve) and  at
$r_{\rm ms}\simeq 1.2M$ (solid curve). Panels (a), (b), and (c),
showing $L_\mp$ vs.~$E_\mp$,
are specific intervals corresponding to Fig.~2$b$ (see text).
\label{fig5}}
\end{figure}                                                                 
\clearpage           

\begin{figure} 
\epsscale{.5}
\plotone{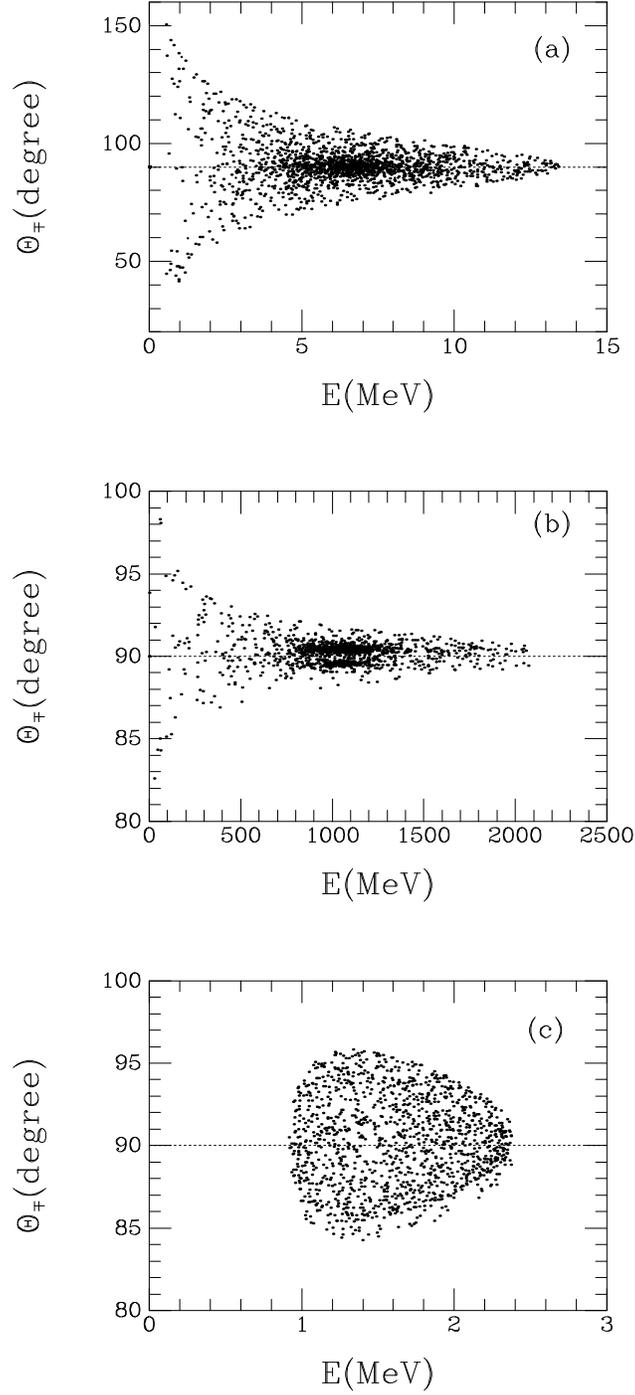}
\caption{Penrose pair production ($\gamma\gamma\lra e^-e^+$):
scatter plots displaying polar
angles, above and below the equatorial
plane ($\Th=90^\circ$), for escaping
$e^-e^+$ pairs (each point
represents an electron).
(a), (b), and~(c), showing $\Th_\mp$ vs.~$E_\mp$,
are the same as cases presented in Figs.~1, 2, and~3,
respectively (see text). \label{fig6}}
\end{figure}
 
\end{document}